\documentclass{JHEP3}
\usepackage{epsfig,amsfonts,amsmath}

\newcommand{\eref}[1]{(\ref{#1})}

\newcommand{\fref}[1]{Figure~\ref{#1}}
\newcommand{\cref}[1]{Chapter~\ref{#1}}
\newcommand{\beq}{\begin{equation}}
\newcommand{\eeq}{\end{equation}}
\newcommand{\ba}{\begin{array}}
\newcommand{\ea}{\end{array}}
\newcommand{\bcenter}{\begin{center}}
\newcommand{\ecenter}{\end{center}}

\def\IC{\mathbb{C}}

\def\IGa{\relax\hbox{${\rm I}\kern-.18em\Gamma$}}

\def\IR{\mathbb{R}}

\def\Z{\mathbb{Z}}
\def\IZ{\mathbb{Z}}

\def\smiley{\hbox{\large$\bigcirc$\hspace{-0.80em}\raise.2ex
\hbox{$\cdot\cdot$}\kern-.61em\lower.2ex\hbox{\scriptsize$\smile$}}\ }
\def\frowny{\hbox{\large$\bigcirc$\hspace{-0.80em}\raise.2ex
\hbox{$\cdot\cdot$}\kern-.635em\lower.2ex\hbox{\scriptsize$\frown$}}\ }

\makeatletter
\let\hangafter\@hangfrom
\makeatother

\newcommand{\be}{\begin{equation}}
\newcommand{\ee}{\end{equation}}
\newcommand{\bea}{\begin{eqnarray}}
\newcommand{\eea}{\end{eqnarray}}
\newcommand{\bean}{\begin{eqnarray*}}
\newcommand{\eean}{\end{eqnarray*}}
\newcommand{\bc}{\begin{center}}
\newcommand{\ec}{\end{center}}
\newcommand{\comment}[1]{}

\newcommand{\beqa}{\begin{eqnarray}}
\newcommand{\eeqa}{\end{eqnarray}}

\newcommand{\bb}{\mathsf{b}}
\newcommand{\ww}{\mathsf{w}}

\preprint{ MIT-CTP-3716\\ 
PUPT-2185}

\title{Moduli Spaces of Gauge Theories from Dimer Models: \\
Proof of the Correspondence}

\author{Sebasti\'an Franco$^1$ and David Vegh$^2$
\\
~\\
$^1$ Joseph Henry Laboratories, \\
Princeton University, \\
Princeton, NJ 08544, USA.
\footnote{
Research supported by the National Science Foundation Grant No. PHY-
0243680.}
\\
\vskip 0.1cm
$^2$ Center for Theoretical Physics, \\
Massachusetts Institute of Technology,\\
Cambridge, MA 02139, USA.\footnote{
Research supported in part by the CTP and the LNS of MIT and the U.S. Department of Energy
under cooperative agreement $\#$DE-FC02-94ER40818.}\\~\\

\email{sfranco@feynman.princeton.edu, dvegh@mit.edu}
}

\abstract{

Recently, a new way of deriving the moduli space of quiver gauge theories that arise on the world--volume of D3--branes probing singular toric Calabi--Yau cones was conjectured.
According to the proposal, the gauge group, matter content and tree--level superpotential of the
gauge theory is encoded in a periodic tiling, the dimer graph. The conjecture provides a simple
procedure for determining the moduli space of the gauge theory in terms of perfect matchings.

For gauge
theories described by periodic quivers that can be embedded on a two--dimensional torus, we prove
the equivalence between the determination of the toric moduli space with a gauged
linear sigma model and the computation of the Newton polygon of the characteristic polynomial of
the dimer model. We show that perfect matchings are in one--to--one correspondence with fields
in the linear sigma model. Furthermore, we prove that the position in the toric diagram of every sigma
model field is given by the slope of the height function of the corresponding perfect matching.

}
\begin{document}

\section{Introduction}

According to the AdS/CFT correspondence \cite{ Maldacena:1997re, Gubser:1998bc, Witten:1998qj,
Aharony:1999ti}, the large $N$ 't Hooft limit of
$\mathcal{N}=4$ $SU(N)$ super Yang Mills is equivalent to type IIB String Theory on $AdS \times
S^5$ with $N$ units of Ramond--Ramond 5--form flux on the $S^5$. The $\mathcal{N}=4$ gauge
theory arises as the worldvolume theory of a stack of $N$ D3--branes in flat ten dimensional
space. Since its original formulation, the AdS/CFT correspondence has been extended to and
checked in a variety of more realistic, less supersymmetric situations. The worldvolume theory
of D3--branes over a singular Calabi--Yau threefold is an $\mathcal{N}=1$ quiver gauge theory
\cite{Douglas:1996sw, Douglas:1997de}. The structure of the gauge theory reflects the properties
of the singular manifold. When the Calabi--Yau is a metric cone over an $X_5$ Sasaki--Einstein
manifold, the corresponding dual is type IIB string theory on $AdS_5 \times X_5$.

Toric Calabi--Yau's are a particularly simple, yet extremely rich, subset in the space of
Calabi--Yau threefolds. Their simplicity resides in that they are defined by a relatively small
amount of combinatorial data and can be constructed in terms of two--dimensional gauged linear
sigma models.

Recently, we have witnessed remarkable progress in our understanding of $\mathcal{N}=1$
superconformal field theories, their embedding in string theory and their AdS/CFT duals. We now
review an abbreviated list of such developments. On the purely field theoretic front, the
a--maximization principle \cite{Intriligator:2003jj} has been a major breakthrough, permitting
the computation of R--charges for arbitrary $\mathcal{N}=1$ superconformal theories. In
\cite{Gauntlett:2004yd, Gauntlett:2004hh}, explicit metrics for an infinite family of
Sasaki--Einstein 5--manifolds denoted $Y^{p,q}$ were found. The metric cones over these
manifolds are toric \cite{Martelli:2004wu} and the corresponding gauge theories have been determined
\cite{Benvenuti:2004dy}. Afterwards, a larger set of metrics dubbed $L^{a,b,c}$, containing the
$Y^{p,q}$'s as particular cases, was discovered \cite{Cvetic:2005ft,Martelli:2005wy,Cvetic:2005vk}. Again, the
corresponding cones are toric and the dual gauge theories were identified \cite{Franco:2005sm,
Benvenuti:2005ja, Butti:2005sw}. With these theories, we passed from having a couple of examples
in which the explicit $AdS_5 \times X_5$ metric and the field theory dual were known ($X_5$
being $S^5$, $T^{1,1}$ and their orbifolds) to an infinite number of such pairs. In
\cite{Martelli:2005tp}, the geometric dual of a--maximization, Z--minimization, was found. Using
Z--minimization it is possible to compute the volume of subcycles in a toric variety using
solely the information in the toric diagram. Further developments in the subject appeared in \cite{Butti:2005vn, Tachikawa:2005tq, Barnes:2005bw}.

In parallel, there has been considerable advancement in the techniques for deriving gauge
theories on D--branes over singularities. Some of the approaches are partial resolution \cite{Morrison:1998cs,Beasley:1999uz, Feng:2000mi} of
orbifold singularities \cite{Douglas:1996sw,Douglas:1997de}, exceptional
collections \cite{Cachazo:2001sg, Wijnholt:2002qz,Herzog:2003dj,Herzog:2003zc,Aspinwall:2004bs, Aspinwall:2005ur, Herzog:2004qw,Herzog:2005sy} and dimer methods \cite{Hanany:2005ve,
Franco:2005rj, Franco:2005sm,Hanany:2005ss,Feng:2005gw,Butti:2005ps}, the subject of this paper. For toric manifolds,
dimers have proved to be very strong in comparison to alternative approaches, producing the most
vast set of results together with an appealing elegance and extreme computational simplicity.

As a result of these developments the paradigm under which we look for and test AdS/CFT pairs,
at least in the case of toric singularities, has changed. Dimer methods immediately provide the
gauge theory for a given toric geometry. Next, we can perform non--trivial checks comparing
R--charges and central charges of the field theory, determined with a--maximization, to the
volumes of supersymmetric cycles in the singular geometry, which can be computed without
explicit knowledge of the metric thanks to Z--minimization.

The dimer method approach to quiver theories on D--branes over toric singularities was initiated
in \cite{Hanany:2005ve}, where a striking correspondence between the perfect matching partition
function and the toric diagram of the underlying geometry was observed. The idea was fully
developed in \cite{Franco:2005rj}, where the rules for constructing
a tiling on which dimers live for an arbitrary toric quiver were established.  A physical interpretation of this tiling as a configuration NS5 and D5--branes was also proposed.
In addition, \cite{Franco:2005rj} conjectured a 
specific correspondence between GLSM fields and perfect matchings, noticing also how perfect matchings
are natural variables to solve F--term equations.
This correspondence, which we call Mathematical Dimer Conjecture in this paper, leads
to impressive simplifications in the study of branes on toric singularities and lies at the core
of the breakthrough of the dimer ideas. The main result of this
paper is the proof of the Mathematical Dimer Conjecture.

Until recently, finding the tiling for a particular theory was somewhat {\it ad hoc}.
A major breakthrough was made in \cite{Hanany:2005ss}, by interpreting R--charges as angles in the
tiling. Furthermore, based on the observation that the so called zig--zag paths are in one--to--one
correspondence with the edges of the toric diagram, the Fast Inverse Algorithm was established
which is by now the most efficient tool for computing the quiver and the superpotential from the
toric diagram of a toric non--compact Calabi--Yau. A physical realization of the tilings, which supports the proposal of \cite{Franco:2005rj}, and a proof
of the Fast Inverse Algorithm was recently derived in \cite{Feng:2005gw} using mirror symmetry.

In order to provide a self--contained presentation, we devote Sections \ref{section_tilings} and
\ref{section_toric} to review background material. Section \ref{section_tilings} discusses the
main concepts in toric quivers, brane tilings and dimer models. Section \ref{section_toric}
presents the gauged linear sigma model (GLSM) approach for computing toric moduli spaces of toric gauge
theories. In Section \ref{section_conjectures} we present the conjecture of
\cite{Franco:2005rj}, splitting it into the Mathematical and Physical dimer conjectures.
Finally, we prove the Mathematical Dimer Conjecture in Section \ref{section_proof}. We
illustrate all discussions in the paper with the relatively non--trivial example of a quiver
theory for D3--branes probing a complex cone over the second del Pezzo surface.

\section{Toric quivers and brane tilings}

\label{section_tilings}

We consider the $\mathcal{N}=1$ superconformal gauge theories that live on the worldvolume of a
stack of $N$ D3--branes probing a non--compact toric Calabi--Yau 3--fold. For every singularity,
the gauge theory on the D3--branes is not unique. In fact, we have an infinite number of gauge
theories connected by Seiberg duality \cite{Seiberg:1994pq, Beasley:2001zp, Feng:2001bn,
Cachazo:2001sg,Franco:2003ja} that flow to the same universality class in the infrared
limit. Every gauge theory is specified by a gauge group and a matter content, which are encoded
in a quiver diagram, and a superpotential. We will concentrate on a particular subset of this
infinite set of dual theories, denoted {\bf toric phases}. A toric phase is defined as a phase
in which the gauge group is $\prod SU(N)$, i.e. the ranks of all gauge group factors are the
same. Non--toric phases are obtained by Seiberg duality on a node for which the number of
flavors is larger than twice the number of colors. The fact that the probed geometry is an
affine toric variety constraints the possible structure of the superpotential. It has to be such
that all F--term equations are of the form {\it ``monomial = monomial''}. This constraint is
dubbed the {\bf toric condition} \cite{Feng:2002zw} and can be rephrased by saying that every
field in the quiver must appear exactly in two terms of the superpotential, with both terms
having opposite signs. In addition, all superpotential coefficients can be normalized to $1$ by
rescaling the fields.

\begin{figure}[ht]
  \epsfxsize = 4cm
  \centerline{\epsfbox{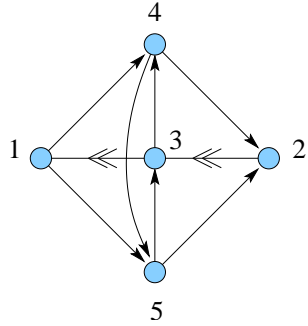}}
  \caption{Quiver diagram for Model II of $dP_2$.}
  \label{quiver_dP2}
\end{figure}

\fref{quiver_dP2} shows one toric phase for the complex cone over $dP_2$ \cite{Feng:2000mi,
Feng:2002zw}, usually referred to as Model II. The corresponding superpotential is given by
\beq
\begin{array}{rl}
W = & [X_{34}X_{45}X_{53}]-[X_{53}Y_{31}X_{15}+X_{34}X_{42}Y_{23}] \\
  + & [Y_{23}X_{31}X_{15}X_{52}+X_{42}X_{23}Y_{31}X_{14}]-[X_{23}X_{31}X_{14}X_{45}X_{52}]
\end{array}
\label{W_dP2}
\eeq
where we have grouped terms to make a $\IZ_2$ global symmetry that acts by interchanging nodes
$1 \leftrightarrow 2$ and $4 \leftrightarrow 5$ and charge conjugating all the fields manifest.
We will use this example to illustrate all our discussions.

In \cite{Franco:2005rj}, it was realized that all the information in the quiver diagram and the
superpotential of a toric phase can be encapsulated in a single object: the {\bf periodic
quiver}. A periodic quiver is a planar quiver drawn on the surface of a 2--torus (equivalently,
a doubly periodic infinite quiver on the plane) s.~t. every plaquette corresponds to a term in
the superpotential. The sign of the superpotential terms is given by the orientation of the
corresponding plaquettes, which alternates between clockwise and counterclockwise. The toric
condition is automatically incorporated in the periodic quiver, since every field appears
precisely in two neighboring plaquettes with opposite orientation.

\begin{figure}[ht]
  \epsfxsize = 8cm
  \centerline{\epsfbox{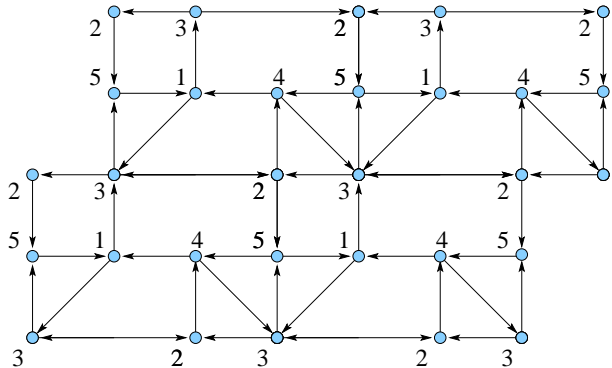}}
  \caption{Periodic quiver for Model II of $dP_2$. We show several fundamental cells.}
  \label{quiver_dP2_II}
\end{figure}

It has been conjectured that any quiver corresponding to D3--branes probing non--compact, toric
Calabi--Yau threefolds can be embedded in a $T^2$  \cite{Franco:2005rj}.
Furthermore, the two cycles around the $T^2$ have been identified with the non--R symmetry
$U(1)$ isometries \cite{Benvenuti:2005cz}. In Section \ref{section_conformal_invariance}, we
discuss how conformal invariance restricts the possible embeddings of the periodic quiver.
\fref{quiver_dP2_II} shows the periodic quiver for our $dP_2$ example.

Along the rest of the paper, our working hypothesis will be that {\it we consider gauge theories
that are described by periodic quivers on $T^2$}. For this class of theories, we will show that the
GLSM determination of the moduli space can be translated into a dimer problem.

The superpotential can be written schematically as
\beq
W=\sum_\mu \pm W_\mu
\eeq
where every superpotential term $W_\mu$ is a gauge invariant mesonic operator with R--charge
equal to 2 and neutral under the $U(1)\times U(1)$ flavor symmetry\footnote{In some cases the
$U(1)^2$ global symmetry can be enhanced. For example, for $Y^{p,q}$ theories the flavor
symmetry is $SU(2)\times U(1)$ \cite{Benvenuti:2004dy}.}. We have explicitly indicated the sign
of each term, which satisfy the toric condition.

In toric quivers, F--term equations can be used to show that all these operators are equivalent
in the chiral ring. The toric condition implies that every field $X_i$ appears (linearly) in
exactly two superpotential terms. Let us call them $W_1$ and $-W_2$ (according to the toric
condition both contributions have opposite signs). Then
\beq
0=X \ \partial_{X} \ W = X \ \partial_{X} \ (W_1-W_2)=W_1-W_2
\eeq

\begin{figure}[ht]
  \epsfxsize = 4cm
  \centerline{\epsfbox{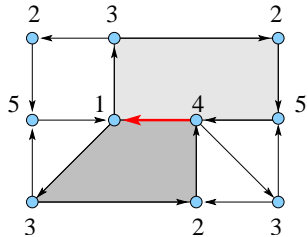}}
  \caption{Two plaquettes are equal once the F--term equation for the   common field is imposed.}
  \label{F_cell_quiver_dP2_II}
\end{figure}

This becomes very intuitive from the perspective of the periodic quiver (see
\fref{F_cell_quiver_dP2_II}), where one can show that any two adjacent plaquettes are equal by
using the F--term relation for the common field. Iterating this process we see that, once
F--term equations are taken into account, all superpotential terms are identical. This idea has
already been used in \cite{Benvenuti:2005cz}.

In \cite{Franco:2005rj}, an alternative representation of the gauge theory, dubbed {\bf brane
tiling}\footnote{We alert the reader that the goal of this paper is independent of the possible
interpretation of the tiling as a physical object, such as a configuration of branes. However,
we will adhere to the term {\bf brane tiling} for simplicity. The arguments that identify brane
tilings with physical configurations of D5-- and NS5--branes are primarily based on the analogy
with brane box and brane diamond constructions dual to orbifold singularities
\cite{Hanany:1997tb, Hanany:1998it, Aganagic:1999fe}. A concrete string theory realization of
the tiling was studied in detail in \cite{Feng:2005gw}.} was introduced. The brane tiling is
constructed by dualizing the periodic quiver graph: Nodes, arrows and plaquettes of the periodic
quiver are replaced by faces, transverse lines and nodes, respectively.

The resulting tiling is a {\bf bipartite graph}. This means that it is possible to assign nodes
two colors (by convention we choose black and white) such that white nodes are only connected to
black nodes and viceversa. The coloring of nodes is in one--to--one correspondence with the
orientation of plaquettes in the periodic quiver (hence the sign of superpotential terms). Edges
in the tiling carry a natural orientation (for example from white to black nodes), which
corresponds to the orientation of arrows in the periodic quiver.

We can translate among periodic quiver, brane tiling and gauge theory concepts using the
following dictionary

\bigskip

\begin{center}
\begin{tabular}{c|c|c}
{\bf Periodic quiver} & {\bf Brane tiling} & {\bf Gauge theory} \\
\hline \hline
node & face & $SU(N)$ gauge group \\
arrow & edge & bifundamental (or adjoint) \\
plaquette & node & superpotential term
\end{tabular}
\end{center}
\bigskip

We denote $F$, $E$ and $N$ the number of faces, edges and nodes in the tiling. They correspond
to the number of gauge groups, chiral multiplets and superpotential terms in the gauge theory.

\begin{figure}[ht]
  \epsfxsize = 7.5cm
  \centerline{\epsfbox{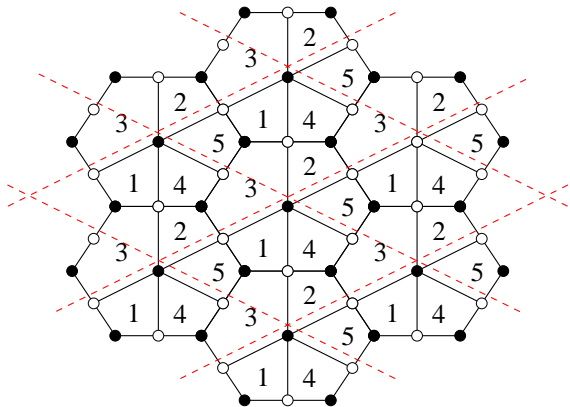}}
  \caption{Brane tiling for Model II of $dP_2$.}
  \label{tiling_dP2_II}
\end{figure}

For a comprehensive description of brane tilings we refer the reader to \cite{Franco:2005rj}.
\fref{tiling_dP2_II} shows the brane tiling for the $dP_2$ example under consideration, obtained
by dualizing the periodic quiver in \fref{quiver_dP2_II}

In analogy to the chemical terminology, every edge in the tiling is denoted a {\bf dimer}. A
{\bf perfect matching} is a collection of edges (dimers) such that every node in the tiling is
the endpoint of exactly one edge in the set. For later reference, we list all perfect matchings
for the $dP_2$ brane tiling in the Appendix. Perfect matchings play a fundamental role in our
forthcoming discussion.

\subsection{Geometry of the tiling embedding from conformal invariance}

\label{section_conformal_invariance}

In the previous section we stated that we will focus on tilings of a two dimensional torus.
Since the gauge theories under consideration have a finite number of gauge groups,
fields and superpotential terms, it is natural to represent them by a tiling of a {\bf compact}
Riemann surface $\Sigma$. But, is any $\Sigma$ a valid option? Why do we choose a $T^2$?
Interestingly, as we discuss in this section, the gauge theory actually constraints
the geometry of $\Sigma$.

Conformal invariance at the IR fixed point requires the beta functions for all superpotential
and gauge couplings to be zero. For superpotential couplings this implies that

\beq
\sum_{{\scriptsize \begin{array}{c} i \in edges \\ around \ node \end{array}}} R_i=2 \ \ \ \ \ \
\mbox{for every node}
\label{beta_nodes}
\eeq
while vanishing of gauge coupling beta functions corresponds to

\beq
2+\sum_{{\scriptsize \begin{array}{c} i \in edges \\ around \ face \end{array}}} (R_i-1)=0 \ \ \ \ \ \
\mbox{for every face}
\label{beta_faces}
\eeq

Adding \eref{beta_faces} over all faces and using \eref{beta_nodes} we conclude that

\beq
F+N-E=\chi(\Sigma)=0
\label{euler}
\eeq

\begin{figure}[ht]
  \epsfxsize = 4cm
  \centerline{\epsfbox{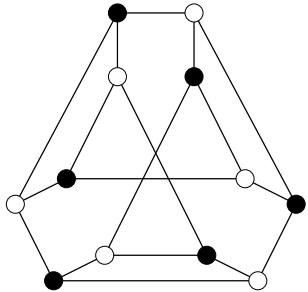}}
  \caption{A bipartite graph tiling the Klein Bottle.}
  \label{tiling_KB}
\end{figure}

Hence, conformal invariance implies that the Euler characteristic of $\Sigma$ has to be zero.
This fact has been already noticed in \cite{Franco:2005rj}. There are only two options for
$\Sigma$. On one hand, it can be a $T^2$ as considered so far in the paper and in the
literature. On the other hand, there is the interesting possibility of $\Sigma$ being the Klein
Bottle. \fref{tiling_KB} shows an example of a bipartite tiling of the Klein Bottle. This tiling
is known as the Franklin graph \cite{bondy} and has $F=6$, $N=12$ and $E=18$, hence satisfying
\eref{euler}.

At present, both the gauge theory and geometric interpretations of such a tiling are
unknown and remain an intriguing question that deserves further study.
Along the rest of the paper, we will restrict ourselves to the case in which $\Sigma=T^2$.
The planar quiver, dual to the tiling, will consequently be also embedded in a $T^2$.

\subsection{Height function}

\label{section_height_function}

Given a perfect matching $M$, it is possible to define an integer--valued {\bf height function}
$h$ over the brane tiling \cite{Kenyon:2002a,Kenyon:2003uj}. In order to do so we fix a
reference perfect matching $M_0$ and a face $f_0$. The difference $M-M_0$ defines a set of
closed curves over the tiling. The minus sign flips the orientation of bifundamentals associated
with the edges of $M_0$, giving the resulting closed curves a definite orientation. The height
function jumps by $\pm 1$ when crossing a curve, where the sign is given by the orientation of
the crossing. The height for $f_0$ is set to be zero. Notice that the difference of the height
functions of two matchings is well--defined independently of $M_0$.

The {\bf slope} of a perfect matching is defined as the height change $(h_x,h_y)$ when moving from
one unit cell to the next one along the two fundamental directions. Changing $M_0$ amounts to a constant
shift $(h_{x0},h_{y0})$ in the slopes of all perfect matchings.

\begin{figure}[ht]
  \epsfxsize = 17cm
  \centerline{\epsfbox{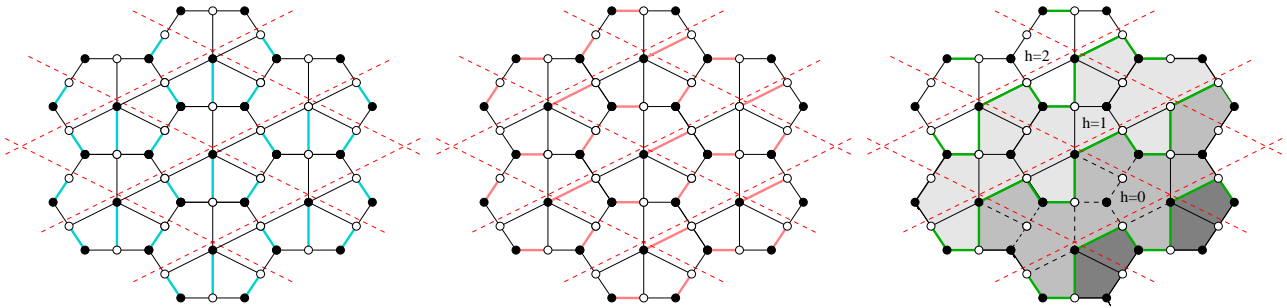}}
  \caption{(a) The dimers in the a perfect matching $M$ are shown in cyan.
  (b) The dimers in the reference perfect matching $M_0$ are shown in red.
  (c) The height function, whose level curves are given by $M-M_0$.}
  \label{height_dP2_2}
\end{figure}

We exemplify the concepts presented in this section with $dP_2$. \fref{height_dP2_2}
shows a perfect matching, a reference perfect matching and the corresponding height function. In this case, we
see that the slope is $(h_x,h_y)=(-1,0)$.

There is an equivalent way to define slopes, that later will turn out to be useful. To every
perfect matching we can associate a {\bf unit flow} on its edges, directed from white to black
nodes. The slope then corresponds to the net flux between adjacent fundamental regions in the $x$ and $y$
directions. The Appendix gives the slopes for all perfect matchings of Model II of $dP_2$. We
will come back to the interpretation of matchings as unit flows in Section \ref{section_pms}.

It is straightforward to count the number of perfect matchings with a given slope \cite{Kenyon:2002a,Kenyon:2003uj}. In order to
do so, we first introduce the Kasteleyn matrix of the tiling $K(x,y)$. It is a weighted, signed,
$(N/2) \times (N/2)$ adjacency matrix defined as follows. In our convention, the rows of
$K(x,y)$ are indexed by white nodes and its columns by black nodes. We associate a $\pm 1$
weight to every edge $e_i$ in the tiling such that when we multiply the weights around every
face we have

\beq
\textrm{sign}\left(\prod e_i \right)=\left\{\begin{array}{l}
+1 \mbox{ if } (\# \mbox{ edges})=2 \mod 4 \\
-1 \mbox{ if } (\# \mbox{ edges})=0 \mod 4
 \end{array}\right.
\eeq

Next we take two fundamental paths $\mathcal{C}_x$ and $\mathcal{C}_y$ in the graph dual to the brane
tiling winding once around the $(1,0)$ and $(0,1)$ cycles of the 2--torus. These paths are
conventionally denoted {\bf flux lines} and can be visualized as the boundaries of the
fundamental region. The weight of every edge in the tiling that is crossed by $\mathcal{C}_x$ is then
multiplied by $x$ or $x^{-1}$ depending on the orientation of the crossing. Respectively, edges
crossed by $\mathcal{C}_y$ are multiplied by $y$ or $y^{-1}$.



The determinant of the Kasteleyn matrix $P(x,y)=\det K(x,y)$ is a Laurent polynomial, the
so--called {\bf characteristic polynomial} of the dimer model. It has the following general form

\beq
P(x,y)=x^{h_{x0}} y^{h_{y0}} \sum c_{h_x,h_y} x^{h_x} y^{h_y}
\eeq

$P(x,y)$ is the partition function of perfect matchings on the brane tiling, in the sense that
the integer coefficients $|c_{h_x,h_y}|$ count the number of perfect matchings with slope
$(h_x,h_y)$ \cite{Kenyon:2003uj}.

In our example, we have

\beq
K=\left( \begin{array}{ccc} 1-x^{-1} & \ \ y \ \ & \ \ 1 \ \ \\
                            1 & 1 & x        \\
                            -1+x^{-1}y^{-1} & 1 & 1 \end{array}\right)
\eeq
Then
\beq
P(x,y)=x^{-1} y^{-1}-x^{-1}+5-x-y-xy
\eeq
which gives the following counting of perfect matchings

{\footnotesize
\begin{center}
\begin{tabular}{|c|c|}
\hline
\ \ \ {\bf slope} \ \ \ & \ \ \ {\bf \# \ matchings} \ \ \ \\
\hline \hline
(-1,-1) & 1 \\
(-1,0)  & 1 \\
(0,0)   & 5 \\
(1,0)   & 1 \\
(0,1)   & 1 \\
(1,1)   & 1 \\
\hline
\end{tabular}
\end{center}
}

\noindent that is in precise agreement with the direct counting in the Appendix.

\section{Toric geometry from gauge theory}

\label{section_toric}

We now review the procedure for computing the moduli space of a given toric quiver (i.e. quiver
plus toric superpotential). For $N$ D3--brane probes, the moduli space along the mesonic branch
corresponds to the symmetric product of $N$ copies of the probed geometry. This procedure has
been algorithmized in \cite{Feng:2000mi} and dubbed the Forward Algorithm. It involves the
following steps:


\begin{itemize}
\item Use F--flatness equations to express the fields in the quiver (which transform
in bifundamental or adjoint representations) $X_i$, $i=1, \ldots ,E$ in terms of $F+2$ independent
variables $v_j$. Although the $v_j$'s can be taken to be a subset of the $X_i$ fields, other
choices are also possible. For example, as we will discuss later, dimers pick other combinations
which turn out to be more natural. The final answer does not depend on this choice. Since for
toric quivers the F--term equations are of the form $monomial=monomial$, each $X_i$ is given by
a product of $v_j$'s to appropriate powers. This can be encoded in an $E \times (F+2)$ matrix
$K$ according to
\beq
X_i=\prod v_j^{K_{ij}}, \ \ \ i=1,\ldots,E, \ \ \ j=1,\ldots, F+2
\eeq
The $X_i$ can involve negative powers of the $v_j$'s, i.e. $K_{ij}$ may have negative entries.
The row vectors $\vec{K}_i$ of $K$ span a cone $M_+$ in $\IR^{F+2}$, corresponding to
non--negative linear combinations of them.

\item Next, to get rid of the negative powers, we introduce new variables $p_\alpha$, $\alpha=1,\ldots,N_{\sigma}$.
In order to do so, we compute the cone $N_+$ dual to $M_+$. $N_+$ is spanned by vectors
$\vec{T}_\alpha$, such that $\vec{K}_i \cdot \vec{T}_\alpha \geq 0$. These vectors can be
organized as the columns of an $(F+2)\times N_\sigma$ integer matrix $T$ such that $K\cdot T
\geq 0$ for all entries. The dimension of the dual cone $N_\sigma$ is not known a priori and is
determined by explicitly computing $N_+$. The intermediate and original variables $v_j$ and
$X_i$ are expressed in terms of the $p_\alpha$ as follow
\beq
v_j=\prod_\alpha p_\alpha^{T_{j\alpha}} \ \ \ \ \ \ \ X_i=\prod_\alpha p_\alpha^{\sum_j K_{ij} T_{j\alpha}}
\eeq
The amount of operations required to compute $N_\sigma$ grows with the size of the gauge theory.
This growth becomes prohibitive when trying to apply the Forward Algorithm to gauge theories
with large quivers. Later, we will explain how this difficulty is circumvented by the dimer
model.

\item A convenient way to encode the relations among the $N_\sigma$ variables $p_\alpha$ and the original $F+2$ $v_j$ is
by obtaining them as D--terms of an appropriately chosen $U(1)^{N_\sigma-(F+2)}$ gauge group.
Its action is given by an $(N_\sigma-F-2)\times N_\sigma$ charge matrix $Q_F$ (where the subindex
$F$ indicates that $Q_F$ contains all the information about F--term equations). Gauge invariance
of the $v_j$'s under the new gauge group gives rise to the desired relations. Hence, $Q_F$ is
such that
\beq
T \cdot Q_F^T=0 \label{T_QF}
\eeq

\item The charges of fields under the $F$ gauge groups of the quiver are summarized by the
$F \times E$ {\bf incidence matrix} $d$. It is defined as
$d_{li}=\delta_{l,head(X_i)}-\delta_{l,tail(X_i)}$. Every column associated to a bifundamental
field contains a $1$ and a $-1$ and the rest of the entries are $0$'s. Adjoint fields are
represented in quiver language by arrows starting from and ending at the same node. Hence, the
corresponding columns have all $0$'s. It is clear that one of the rows of $d$ is redundant.
Thus, we define the matrix $(F-1)\times E$ matrix $\Delta$, which is obtained from $d$ by
deleting one of its rows. For our example, we have

{\footnotesize
\beq
\Delta=\left[\begin{array}{c|ccccccccccc}

\ \ \ \ & \ X_{14} \ & \ X_{31} \ & \ X_{15} \ & \ Y_{31} \ & \ X_{23} \ & \ X_{52} \ & \ Y_{23} \ & \ X_{42} \ & \ X_{34} \ & \ X_{53} \ & \ X_{45} \ \\
\hline
1 & -1 & 1 & -1 & 1 & 0 & 0 & 0 & 0 & 0 & 0 & 0 \\
2 & 0 & 0 & 0 & 0 & -1 & 1 & -1 & 1 & 0 & 0 & 0 \\
3 & 0 & -1 & 0 & -1 & 1 & 0 & 1 & 0 & -1 & 1 & 0 \\
4 & 1 & 0 & 0 & 0 & 0 & 0 & 0 & -1 & 1 & 0 & -1
\end{array}
 \right]
\label{delta}
\eeq
}

The $F-1$ independent D--term equations of the original theory are implemented by adding
a $U(1)^{F-1}$ gauge symmetry to the GLSM. The charges of the $p_\alpha$ under these symmetries
is given by an $(F-1)\times N_\sigma$ matrix $Q_D$ which can be determined in two steps. First,
we construct an $(F-1)\times (F+2)$ matrix $V$ that translates the charges of the $X_i$'s to
those of the $v_j$'s. Thus,
\beq
V \cdot K^T=\Delta \label{VK}
\eeq

Next, we find an $(F+2)\times N_\sigma$ matrix $U$ that transform the charges of $v_j$'s into
those of the $p_\alpha$'s
\beq
U \cdot T^T=\mbox{Id}_{(F+2)\times (F+2)} \label{UT}
\eeq

Finally, we have
\beq
Q_D=V \cdot U \label{QD}
\eeq

$Q_D$ and $Q_F$ are combined into a single $(N_\sigma-2)\times N_\sigma$ charge matrix $Q$
\beq
Q=\left(\begin{array}{c} Q_D \\ Q_F \end{array}\right)
\eeq

The construction we outlined can interpreted as a Witten's two dimensional {\bf gauged linear sigma model} (GLSM) of
$N_\sigma$ chiral fields $p_\alpha$ and $U(1)^{N_\sigma-3}$ gauge group with charges given by $Q$.

\item The $U(1)$ charges defined above are exactly those that appear in the construction of a toric variety as
a symplectic quotient. In toric geometry it is standard to encode the charge matrix by means of
a {\bf toric diagram}.
\beq
G=\left(\mbox{Ker}(Q)\right)^T \label{G_ker_Q}
\eeq
One of the rows in $G$ can be set to have all entries equal to $1$ by an appropriate $SL(3,\IZ)$
transformation. This is the Calabi--Yau condition and amounts to the fact that the sum of the
charges of all the $p_\alpha$ under any of the $U(1)$ gauge symmetries is zero. Effectively, we
are left with a two dimensional toric diagram. Every GLSM field $p_\alpha$ corresponds to a
point in the toric diagram, which is a vector $\vec{v}_\alpha$ in $\IZ^3$. $Q$ is given by
linear relations of the form
\beq
\sum_{i=1}^n Q_a^\alpha \vec{v}_\alpha=0
\eeq
satisfied by the $\vec{v}_\alpha$'s.

\end{itemize}

\begin{figure}[ht]
\begin{center}
  \epsfxsize = 7cm
  \centerline{\epsfbox{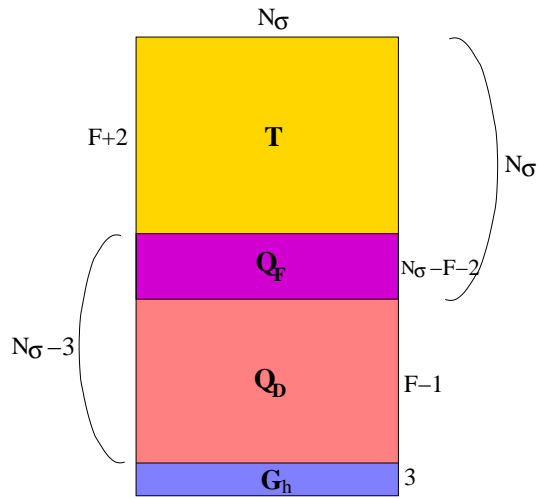}}
  \caption{Relevant matrices in the Forward Algorithm.}
  \label{matrices}
\end{center}
\end{figure}

\bigskip
\fref{matrices} summarizes the relevant matrices in the Forward Algorithm.

\section{The conjecture}

\label{section_conjectures}

Having introduced all necessary concepts, we are ready to study the conjecture of
\cite{Franco:2005rj}. It is convenient to divide the conjecture into two parts, to which we
refer as the {\bf Mathematical} and the {\bf Physical Dimer Conjectures}.

\subsection*{Mathematical dimer conjecture}

The mathematical dimer conjecture states that there is a one--to--one correspondence between
fields $p_\alpha$ in the gauged linear sigma model construction of the toric moduli space of the
given toric gauge theory and perfect matchings in the brane tiling dual to the toric quiver.
Here, when we refer to a toric gauge theory we mean a gauge theory whose quiver can be drawn on a surface of a 2--torus, s.~t. the plaquettes
give the terms in the superpotential (see discussion in Section \ref{section_conformal_invariance}). Furthermore, according to the conjecture, the toric
diagram is the Newton polygon of the characteristic polynomial (i.e. the set of integer exponents of monomials \cite{Kenyon:2003uj}) which, as we have already discussed, is the
set of height function monodromies (``slopes'') of the perfect matchings.

\subsection*{Physical dimer conjecture}

The physical dimer conjecture identifies dimers and tilings with physical objects. According to
the conjecture, the brane tiling is interpreted as a physical brane configuration. It consists
of an NS5--brane extended in the $0123$ directions that wraps an holomorphic curve in $4567$.
The $5$ and $6$ directions are periodically identified giving rise to the 2--torus. D5--branes
extend in $012346$, suspended within the ``holes'' of the NS5--brane in the $46$ torus. Every
stack of D5--branes gives rise to a gauge group. Strings crossing every NS5--brane segment and
connecting two D5--brane stacks correspond to chiral multiplets transforming in the
bifundamental representation of the corresponding gauge groups. Gauge invariant superpotential
terms are produced by the coupling of massless string states at the nodes of the NS5--brane
configuration. This configuration is conjectured to be related to the D3--branes over the
singularity by two T--dualities. The suspended D5--branes are dual to the probe D3--branes and
the NS5--brane structure is dual to the singular geometry.


The correspondence between dimers and a physical brane system could be more subtle and might
differ from the one suggested by the physical dimer conjecture. However, the validity of the
mathematical dimer conjecture, which is the main subject of this paper, is completely
independent of how tilings are realized in terms of branes\footnote{Recently, another physical
description of the tiling has been developed in \cite{Feng:2005gw}. Using mirror symmetry, the D3--branes are
mapped to a system of D6--branes that wraps a self--intersecting $T^3$ torus. The mirror
geometry is a double fibration over the complex $W$ plane, one being the $W=uv$ torus fibration
degenerating at the origin and another being the $W=P(w,z)$ fibration degenerating at some
critical points. Here $P(w,z) \equiv \mbox{det(Kasteleyn)}$ is the spectral curve with $
(w,z)=(e^{s+i \theta}, e^{t+i \phi}) \in (\IC^{\ast} )^2 $. The spectral curve can then be
projected to the non--compact space $(s,t)$ which yields the amoeba whose spine is the pq--web of
the toric diagram. Projection on the compact $(\theta, \phi)$ coordinates gives the so--called
alga of the curve. Its skeleton is the rhombus loop diagram that has been used to construct the
brane tiling for a given toric diagram \cite{Hanany:2005ss, Feng:2005gw}. This construction supports the D5--NS5 tiling proposal 
of \cite{Franco:2005rj}, which appears when T--dualizing along the $S^1$ fibre in the $uv$ plane.}.


\bigskip

Having introduced the conjectures of \cite{Franco:2005rj}, we devote the rest of the paper to
proving the mathematical dimer conjecture.

\section{The proof}

\label{section_proof}

In this section we prove the Mathematical Dimer Conjecture. As we said before, we prove it for
toric gauge theories whose quivers (and hence their brane tilings) are embedded in a two--torus.
A considerable amount of evidence supporting its validity has been accumulated in the
literature. This includes:

\begin{itemize}
\item Construction of the correct toric diagram for the moduli space of gauge theories for an infinite number of singularities. This number is infinite thanks to the
determination of the tilings for the $Y^{p,q}$ \cite{Franco:2005rj} and $L^{a,b,c}$ manifolds
\cite{Franco:2005sm,Butti:2005sw}.
\item Precise agreement between the number of perfect matchings and the multiplicity of
GLSM fields in toric diagrams for various models \cite{Franco:2005sm}.
\item Derivation of Seiberg dual theories by transformations of the tilings preserving
the Newton polygon of the characteristic polynomial \cite{Franco:2005sm,
Hanany:2005ss}.
\item In \cite{Franco:2005sm}, it was shown that given a simple proposal to express quiver fields in terms
of perfect matchings, F--term conditions are straightforwardly satisfied. This proposal will be
derived as part of our proof.
\item The geometry of brane tilings has recently been investigated in \cite{Feng:2005gw}.
The results of this paper show how tilings appear in the description of toric gauge theories by explicitly deriving them from the mirror geometry but do not prove the correspondence between perfect matchings and GLSM fields.

\end{itemize}

Our computations with dimers will closely follow those of the Forward Algorithm. It is important
to keep in mind that some of the steps (or intermediate matrices) are naturally skipped by the
inherent simplifications of the dimer approach. In order to avoid confusion we will use tilded
variables at some stages of the proof. In the end, we will show that they can be identified with
the untilded ones of the Forward Algorithm.

\subsection{Solving F--term conditions: gauge transformations and magnetic coordinates}

\label{section_Fterm_gauge}

The tiling is bipartite, therefore each edge has a natural orientation from its white vertex to
its black vertex. Any weight function $\epsilon(e)$ on the edges defines a 1--form, satisfying
$\epsilon(-e)=-\epsilon(e)$, where $-e$ is the edge with opposite direction
\cite{Kenyon:2003uj}. We denote the linear space of 1--forms on the tiling by $\Omega^1$.
Analogously, the functions on nodes and faces define 0-- and 2--forms in $\Omega^0$ and
$\Omega^2$. The three spaces are related by differentials

\beq
0 \rightarrow \Omega^0 \xrightarrow{\,d\,} \Omega^1 \xrightarrow{\,d\,} \Omega^2 \rightarrow 0
\eeq

We can now define {\bf gauge transformations} on the tiling, whose action on the 1--forms is given by \cite{Kenyon:2003uj}
\beq
\epsilon'(e_i)=\epsilon(e_i)+df \ \ \ \ \ \ \ f \in \Omega^0
\eeq

That is
\beq
\epsilon'(e_i)=\epsilon(e_i)+f(\bb_i)-f(\ww_i)
\eeq
with $\bb_i$ and $\ww_i$ the black and white nodes at the endpoints of
edge $e_i$. These gauge transformations of the tiling should not be confused with the gauge symmetries of the quiver theory. We are confident that the distinction between both types of gauge transformations will be clear from the context in which we use them.
Given a closed path on the tiling
\beq
\gamma=\{\ww_0,\bb_0,\ww_1,\bb_1,\ldots,\bb_{k-1},\ww_k\} \ \ \ \ \ \ \ww_k=\ww_0
\eeq
we define the {\bf magnetic flux} through $\gamma$ as
\beq
B(\gamma)=\int_\gamma \epsilon=\sum_{i=1}^{k-1}\left[\epsilon(\ww_i,\bb_i)- \epsilon(\ww_{i+1},\bb_i)\right]
\eeq

Magnetic fluxes are clearly gauge invariant. The brane tiling is embedded in a two dimensional
torus. Hence, gauge inequivalent classes of 1--forms are parameterized by $\IR^{F-1}\oplus
\IR^2$. The first term corresponds to $d\epsilon \in \Omega^2$, a function on the faces of the
tiling subject to the condition $\sum d\epsilon=0$. We can specify the $\IR^{F-1}$ part by the
magnetic fluxes $B_z(j)$ ($j=1,\ldots,F-1$) through the $\gamma_i$ contours around $F-1$ faces.
The remaining two parameters $(B_x,B_y)$ correspond to fluxes around two non--trivial cycles
($\gamma_x, \gamma_y$) winding around the torus.

Gauge transformations are of particular interest because taking $\epsilon$ to be the energy
function they do not modify the energy difference between perfect matchings. Hence, the
probability distribution of perfect matchings is invariant under gauge transformations.

In this section, we will exploit gauge transformations with a different goal, namely to provide
a convenient set of variables (mostly in $\Omega^2$) that solve the F--term equations. For this
purpose, we define the complex 1--form
\beq
\epsilon(e_i) = \ln X_i \ \ \  \Rightarrow \ \ \ \mbox{under gauge transformations: } X_i'=X_i
e^{f(\bb_i)-f(\ww_i)} \label{gauge_Xi}
\eeq
In this context, we refer to the $X_i$'s as {\bf weights}\footnote{If we regard $-\epsilon(e_i)$
as the energy of a link, the $X_i$'s can be interpreted as complex valued Boltzmann weights.}.

Using \eref{gauge_Xi}, we can define new variables associated to closed paths
\beq
\tilde{v}(\gamma)=e^{\int_\gamma \epsilon}=\prod_{i=1}^{k-1} {X(\ww_i,\bb_i)\over X(\ww_{i+1},\bb_i)}
\label{flux_v}
\eeq
where the product runs over the contour $\gamma$. Then $\{ \tilde{v}_j \equiv
\tilde{v}(\gamma_j), \tilde{v}_x, \tilde{v}_y \}$ provides a parametrization of inequivalent
gauge classes.

We define a convenient basis of 0--forms $F^{(\mu)}$, $\mu=1,\ldots,N$,
\beq
F^{(\mu)} \ \
\left\{ \begin{array}{l} f_\mu=1  \\
                         f_\nu=0 \mbox{ for } \nu \neq \mu
\end{array} \right.
\label{basic_0-forms}
\eeq

Their virtue is that superpotential terms transform simply under the corresponding gauge transformations.
Taking the gauge transformation for $\alpha_\mu F^{(\mu)}$, with $\alpha_\mu$ a complex coefficient, we get
\beq
W_\mu'=W_\mu e^{\textrm{sign}(\mu) v_\mu \alpha_\mu} \label{basic_gauge_transformations}
\eeq
where $v_\mu$ is the valence of node $\mu$ (i.e. the order of the associated superpotential term
$W_\mu$) and following \eref{gauge_Xi} $\textrm{sign}(\mu)$ is $1$ for black nodes and $-1$ for
white nodes.

As discussed in Section \ref{section_tilings}, solving F--term conditions corresponds to setting
all the $W_\mu$'s equal. Given arbitrary values of the $W_\mu$, it is possible to set them equal
to $W_1$ by the basic gauge transformations of \eref{basic_gauge_transformations} with
\beq
\alpha_\mu={\textrm{sign}(\mu)\over v_\mu} {\ln W_1 \over \ln W_\mu}
\eeq

In other words, solving F--term equations corresponds in this language to {\bf partially fixing
the gauge}. Each gauge choice can be labeled by the common value of $W_\mu=W_1$ \footnote{We thank Alastair King for discussions on related ideas.}. Equivalently, one
can label gauge choices using the more symmetric variable $\mathcal{V}$ defined as
\beq
\mathcal{V}=W_1^N=\prod_{\mu=1}^N W_\mu=\prod_{i=1}^E X_i^2 
\eeq
We denote $\mathcal{V}$, the $\tilde{v}_j$'s, $\tilde{v}_x$ and $\tilde{v}_y$ the {\bf flux
variables}.

We have just seen that on each gauge orbit there exists a unique solution to F--term equations
for every value of $\mathcal{V}$. Hence, we conclude that {\it solutions to F--flatness
equations are parametrized by the $F+2$ flux variables: the value of $\mathcal{V}$ indicating a
partial gauge fixing, along with the variables $\tilde{v}_j$ ($j=1,\ldots, F-1$), $\tilde{v}_x$
and $\tilde{v}_y$ parametrizing gauge equivalence classes.} It is now clear that these fluxes
can be identified with the $v_j$ ($j=1,\ldots,F+2$) variables of the Forward Algorithm.

With this identification, it is straightforward to write down a left inverse matrix for $K$, which we
call $K_L^{-1}$. This is an $(F+2)\times E$ matrix such that $K_L^{-1} K=\mbox{Id}_{(F+2)\times
(F+2)}$.


For our $dP_2$ example, we have

\begin{figure}[ht]
  \epsfxsize = 9cm
  \centerline{\epsfbox{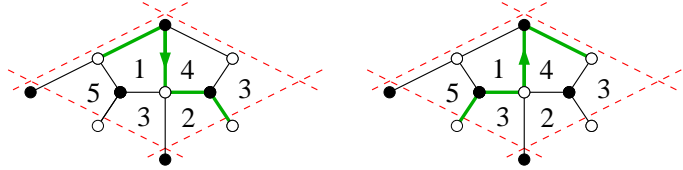}}
  \caption{Contours defining $\tilde{v}_x$ and $\tilde{v}_y$.}
  \label{gammaxy_dP2_II}
\end{figure}

{\footnotesize
\beq
K_L^{-1}=\left[\begin{array}{c|ccccccccccc}

\ \ \ \ & \ X_{14} \ & \ X_{31} \ & \ X_{15} \ & \ Y_{31} \ & \ X_{23} \ & \ X_{52} \ & \ Y_{23} \ & \ X_{42} \ & \ X_{34} \ & \ X_{53} \ & \ X_{45} \ \\
\hline
\tilde{v}_1 & 1 & -1 & 1 & -1 & 0 & 0 & 0 & 0 & 0 & 0 & 0 \\
\tilde{v}_2 & 0 & 0 & 0 & 0 & 1 & -1 & 1 & -1 & 0 & 0 & 0 \\
\tilde{v}_3 & 0 & 1 & 0 & 1 & -1 & 0 & -1 & 0 & 1 & -1 & 0 \\
\tilde{v}_4 & -1 & 0 & 0 & 0 & 0 & 0 & 0 & 1 & -1 & 0 & 1 \\
\hline
\tilde{v}_x & -1 & 0 & 0 & 1 & -1 & 0 & 0 & 1 & 0 & 0 & 0 \\
\tilde{v}_y & 1 & -1 & 0 & 0 & 0 & 0 & 0 & 0 & 0 & 1 & -1 \\
\hline
\mathcal{V} & 2 & 2 & 2 & 2 & 2 & 2 & 2 & 2 & 2 & 2 & 2
\end{array}
 \right]
\eeq
} for which we have taken the $\gamma_i$ loops to run clockwise around faces, and $\gamma_x$ and
$\gamma_y$ are the two non--trivial cycles shown in \fref{gammaxy_dP2_II}, i.e.

\beq
\begin{array}{l}
\tilde{v}_x=X_{14}^{-1} X_{42} X_{23}^{-1} Y_{31} \\
\tilde{v}_y=X_{53} X_{31}^{-1} X_{14} X_{45}^{-1}
\end{array}
\eeq

With this choice of contours, it is clear that the first $F-1$ rows of $K_L^{-1}$ are equal to
$-\Delta$ (see \eref{delta}). There are other paths equivalent to $\gamma_x$ and $\gamma_y$ that
are obtained by deforming them using F--term equations.

The matrix $K$ converts magnetic variables into weight variables. We do not determine $K$
explicitly in this section as it is not necessary for our discussion. As explained in Section
\ref{section_toric}, the vectors $\vec{n}_i$ corresponding to rows in $K$ ($i=1,\ldots, E$) span
a cone $\mathcal{S}$ in $\IR^{F+2}$.

\subsection{The GLSM fields are perfect matchings}
\label{section_pms}

In the previous section we discussed at length how the F--flatness conditions can be satisfied
in terms of the $\tilde{v}_i$ magnetic fluxes that are in one--to--one correspondence with the
variables $v_j$ according to \eref{flux_v}.
The relation between these variables and the original $X_i$ fields are encoded in the matrix $K$, whose rows span the cone $M_+$ in $\IR^{F+2}$. The Forward Algorithm proceeds by computing the
cone dual to $M_+$:
\be
  N_+=\{x \in \IR^{F+2} \ | \ \langle \vec{K}_i , x \rangle \ge 0 \ \ \mbox{for} \ i=0,\dots ,E \}
\ee
There are $N_{\sigma}$ spanning vectors for this dual cone $N_+$. These $N_{\sigma}$ vectors
define the columns $\vec{T}_j$ of the matrix $T$ and they are in one--to--one correspondence
with the homogeneous $p_{\alpha}$ GLSM coordinates.

We would like to understand the computation of the dual cone in terms of tiling techniques. In
order to do so, we introduce a slightly different viewpoint that will prove to be useful.

An arbitrary real weight system on the edges can be interpreted as a {\bf white--to--black
flow}\footnote{The flow space should not be confused with the flux space, which was introduced
in the previous section and is $\IR^{F+2}$.} \cite{Kenyon:2003uj}. The (possibly negative) strength of the flow from
white to black node along an edge $e_i$ is given by the corresponding real weight $c_i$. The
real weights considered in this section are not to be confused with the complex weights given by
$X_i$ that we have discussed earlier.

A flow is nonnegative if it has a nonnegative strength on all edges of the tiling ($c_i >
0$~for~all~$e_i$). The flows are typically not divergence free, therefore there can be sinks and
sources at the vertices. The net flux coming out of a given white node or into a black node is denoted the 
{\bf vorticity} of the node. 

For each point in flux space $x \in \IR^{F+2}$ we define a real flow on the tiling whose
strength at the $i^{\textrm{th}}$ edge is given by $\sum_{j} K_{ij} x_{j}$. Hence the points
inside $N_+$ correspond to nonnegative flows in this picture.

We want to find the spanning vectors $\vec{T}_{\alpha}$ of the dual cone $N_+ \in \IR^{N_F+2}$. Following our discussion in Section \ref{section_Fterm_gauge}, 
we can rescale the vectors $\vec{T}_{\alpha}$ by a positive real number using the gauge transformations of the dimer model. 
Thus we can set their vorticity to one. Therefore, we can focus on the hyperplane $H \subset \IR^{N_F+2}$ such 
there is a unit source residing at every white vertex and a unit sink at the black ones. 
The flows associated with this hyperplane are called {\bf unit flows}. 


The vectors $\vec{T}_{\alpha}$ span the cone $N_+$, hence they also span the intersection $H\cap
N_+$ in flux space. From the previous discussion, we know that this intersection is linearly
mapped by $K_{ij}$ to nonnegative unit flows $P \subset \IR^{E}$ in flow space. It is
well--known in the literature that the set of nonnegative unit flows is a convex polytope in the
flow space and that perfect matchings are vertices of this polytope (Perfect Matching Polytope
Theorem, \cite{Edmonds:1965}). Their preimages are the spanning vectors $\vec{T}_{\alpha}$ in
flux space. For $\vec{T}_{\alpha}$, the flow on the $i^{\textrm{th}}$ edge is given by $\sum_{j}
K_{ij} (\vec{T}_{\alpha})_{j} = \sum_{j} K_{ij} T_{j\alpha}$. We conclude that there is a
one--to--one correspondence between the GLSM fields in the Forward Algorithm and perfect
matchings.

Perfect matchings are naturally represented as unit flows, hence they immediately determine
$KT$. By introducing the following ``product'' between perfect matchings and edges in the tiling
\beq
\langle e_i,p_\alpha\rangle=\left\{ \begin{array}{l} 1 \mbox{ if } e_i \in p_\alpha \\
                                        0 \mbox{ if } e_i \notin p_\alpha  \end{array} \right.
\label{projection_pm}
\eeq
the matrix $KT$ is simply
\beq
(KT)_{i\alpha} = \langle e_i,p_\alpha \rangle 
\label{KT_product}
\eeq
The correspondence between GLSM fields and perfect matchings and the computation of $KT$ in terms of perfect matchings
that we derived in this section was originally proposed in \cite{Franco:2005rj}.

Using \eref{KT_product} for $dP_2$, we have {\footnotesize
\beq
KT^T=\left[\begin{array}{c|ccccccccccc}

\ \ \ \ & \ X_{14} \ & \ X_{31} \ & \ X_{15} \ & \ Y_{31} \ & \ X_{23} \ & \ X_{52} \ & \ Y_{23} \ & \ X_{42} \ & \ X_{34} \ & \ X_{53} \ & \ X_{45} \ \\
\hline
p_1 & 0 & 1 & 0 & 1 & 0 & 0 & 0 & 0 & 1 & 0 & 0 \\
p_2 & 0 & 0 & 1 & 0 & 0 & 0 & 0 & 1 & 0 & 0 & 1 \\
p_3 & 0 & 0 & 0 & 0 & 0 & 1 & 0 & 1 & 0 & 1 & 0 \\
p_4 & 0 & 0 & 0 & 0 & 1 & 0 & 1 & 0 & 0 & 1 & 0 \\
p_5 & 1 & 0 & 1 & 0 & 0 & 0 & 0 & 0 & 1 & 0 & 0 \\
p_6 & 0 & 0 & 0 & 1 & 0 & 0 & 0 & 1 & 0 & 1 & 0 \\
p_7 & 1 & 0 & 0 & 0 & 1 & 0 & 0 & 0 & 0 & 1 & 0 \\
p_8 & 0 & 1 & 0 & 0 & 1 & 0 & 0 & 0 & 0 & 0 & 1 \\
p_9 & 0 & 0 & 1 & 0 & 0 & 0 & 1 & 0 & 1 & 0 & 0 \\
p_{10} & 0 & 1 & 0 & 0 & 0 & 1 & 0 & 0 & 1 & 0 & 0
\end{array}\right]
\eeq
}

As we discussed in the previous section, the left inverse of $K$, which we called $K_L^{-1}$,
arises naturally using dimer methods. Then, it is straightforward to write down
\beq T=K_L^{-1} K
T
\eeq

{\footnotesize
\beq
T=\left[\begin{array}{c|cccccccccc}

\ \ \ \ & \ p_1 \ & \ p_2 \ & \ p_3 \ & \ p_4 \ & \ p_5 \ & \ p_6 \ & \ p_7 \ & \ p_8 \ & \ p_9 \ & \ p_{10} \  \\
\hline
\tilde{v}_1 & -2 & 1 & 0 & 0 & 2 & -1 & 1 & -1 & 1 & -1 \\
\tilde{v}_2 & 0 & -1 & -2 & 2 & 0 & -1 & 1 & 1 & 1 & -1 \\
\tilde{v}_3 & 3 & 0 & -1 & -3 & 1 & 0 & -2 & 0 & 0 & 2 \\
\tilde{v}_4 & -1 & 2 & 1 & 0 & -2 & 1 & -1 & 1 & -1 & -1 \\
\hline
\tilde{v}_x & 1 & 1 & 1 & -1 & -1 & 2 & -2 & -1 & 0 & 0 \\
\tilde{v}_y & -1 & -1 & 1 & 1 & 1 & 1 & 2 & -2 & 0 & -1 \\
\hline
\mathcal{V} & 6 & 6 & 6 & 6 & 6 & 6 & 6 & 6 & 6 & 6
\end{array}\right]
\eeq
}

Notice that the fact that $T$ may have negative entries is not a problem. The important point is
that $(KT)_{i\alpha}\geq 0$. In fact we can give a straightforward definition of $T$ in terms of
the tiling, similar to \eref{KT_product}. In order to do so, we take into account the edges $e_i$ in
the curves $\gamma_j$ that define the magnetic fluxes (similarly, all $e_i$'s are included for $\mathcal{V}$). The $\gamma_j$'s have an {\bf orientation} and then the fields $X_i$
associated to edges $e_i$ appear with a $\pm 1$ power that we denote $\textrm{sign}(e_i)$.
Combining these ideas, we get
\beq
T_{j \alpha}=\sum_{e_i \in \gamma_j} \textrm{sign}(e_i) \langle e_i,p_\alpha \rangle
\eeq

\subsection{Height changes as positions in a toric diagram}

So far we have shown that GLSM fields are perfect matchings. This is half of the proof
of the Mathematical Dimer Conjecture, which in addition states that the height changes
$(h_x,h_y)$ of a given perfect matching should be interpreted as the position in the
toric diagram of the corresponding GLSM field.

Let us define the following $3\times N_\sigma$ matrix
\beq
G_h=\left(\begin{array}{c} h_x \\ h_y \\ 1 \end{array}\right)
\label{Gh}
\eeq

The non--trivial piece of $G_h$ is given by $(h_x,h_y)$. We have included a third row with value
$1$ for all perfect matchings that plays the role of the trivial coordinate of the toric
diagram.

Our goal is to prove that $G_h$ defines the GLSM charge matrix $Q$ through the vanishing linear
relations among its columns, and thus can be identified with $G$
in \eref{G_ker_Q}. I.e. we want
to show that
\beq
\begin{array}{ccr}
Q \ G_h^T=0 & \ \ \ \ \Leftrightarrow \ \ \ \ & Q_F \ G_h^T=0 \\
            &                                 & \mbox{and  } Q_D \ G_h^T=0
\end{array}
\label{condition_Gh}
\eeq

For the third row of $G_h$, \eref{condition_Gh} means that the trace over perfect
matchings of any given GLSM $U(1)$ charge vanishes. It is straightforward to see that this
condition is always satisfied. Thus, from now on we concentrate on
the $(h_x,h_y)$ piece of $G_h.$

Let us first show that $Q_F \ G_h^T=0$. From \eref{T_QF}, we have
\beq
T \ Q_F^T=0
\label{T_QF}
\eeq

Hence, it is sufficient to prove that $h_x$ and $h_y$ are given by linear combinations of the rows of $T$.
It is straightforward not only to show that this is the case but also to identify the precise form of these
linear combinations. The key ideas are the interpretation of height changes as horizontal and vertical net flows
as discussed in Section \ref{section_height_function} and that $KT$ is computed as the ``overlap'' of perfect matchings and edges \eref{KT_product}.
With this in mind, we can express the height changes as
\beq
h_x(p_\alpha)=\sum_j ( \sum_{e_i \in E_x} \textrm{sign}^{x}(e_i) K_{ij} ) \ T_{j\alpha}
\label{linear_h_T1}
\eeq

\beq
h_y(p_\alpha)=\sum_j ( \sum_{e_i \in E_y}  \textrm{sign}^{y}(e_i) K_{ij} ) \ T_{j\alpha}
\label{linear_h_T2}
\eeq
where $E_x$ and $E_y$ denote the set of edges crossing the horizontal and vertical boundaries of
the unit cell (i.e. the flux lines $\mathcal{C}_x$ and $\mathcal{C}_y$), and $\textrm{sign}^{x}(e_i)$ and
$\textrm{sign}^{y}(e_i)$ indicate the direction of the crossing. For illustration, let us
consider our $dP_2$ example, for which
\beq
\begin{array}{lcl}
E_x=\{X_{52},X_{53},Y_{23}\} & \ \ \ \ \ \ \ \ & \textrm{sign}^{x}(e_i)=\{-1,1,-1 \} \\
E_y=\{X_{23},Y_{23}\}        & \ \ \ \ \ \ \ \ & \textrm{sign}^{y}(e_i)=\{1,-1 \}
\end{array}
\eeq

\begin{figure}[ht]
  \epsfxsize = 9cm
  \centerline{\epsfbox{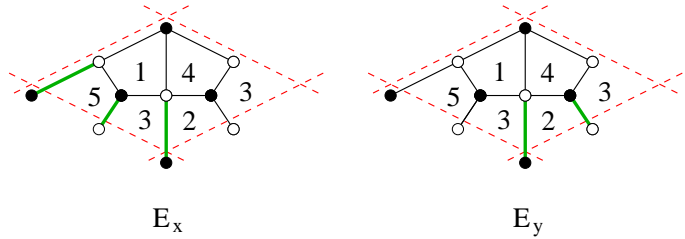}}
  \caption{Sets of edges $E_x$ and $E_y$ that enter the computation of $(h_x,h_y)$.}
  \label{Ex_Ey}
\end{figure}

\fref{Ex_Ey} shows $E_x$ and $E_y$ in the tiling.

Using \eref{T_QF}, the fact that $(h_x,h_y)$ is given by the linear combinations constructed in
\eref{linear_h_T1} and \eref{linear_h_T2} implies that
\beq
Q_F \ G_h^T=0
\eeq
as we want. The missing part of the proof is to show that $Q_D \ G_h^T=0$. This can be done as follows

\begin{eqnarray}
\left( Q_D \ G_h^T \right)_{l x} & = & \sum_{e_i \in E_x} \textrm{sign}^{x}(e_i) \left(VUT^TK^T
\right)_{li}=
\sum_{e_i \in E_x} \textrm{sign}^{x}(e_i) \left(VK^T \right)_{li} \nonumber \\
                                 & = & \sum_{e_i \in E_x} \textrm{sign}^{x}(e_i) \Delta_{li}=0
\label{QD_hx}
\end{eqnarray}
In the first equality we have used \eref{linear_h_T1} and \eref{QD}. In the second one, we used
\eref{UT}. In the third one, we used \eref{VK}. The last step uses the following reasoning.
Every face $l$ of a tiling ($l=1,\ldots, F$) is crossed by $\mathcal{C}_x$ over an even number of
edges \footnote{Actually, a face of the tiling may be crossed by $\mathcal{C}_x$ over an odd number
of edges. This happens when there are chiral multiplets transforming in the adjoint
representation of the corresponding gauge group. Adjoint fields are
represented in the tiling by edges such that the faces at both of its sides are identified (arrows beginning and ending at the same node in the dual quiver). For
a field $X_i$ in the adjoint representation of the $l^{\textrm{th}}$ gauge group $\Delta_{li}=0$
and thus the derivation of \eref{QD_hx} still holds. The reader should keep in mind this
subtlety.}. Typically, as in the $dP_2$ example we are considering in the paper, this
intersection number is 0 or 2, but larger values are also possible. Every edge intersected by
$\mathcal{C}_x$ corresponds to a field $X_i$ in $E_x$ that transforms either in the fundamental
($\Delta_{li}=1$) or antifundamental ($\Delta_{li}=-1$) representation of the $SU(N)$ gauge
group associated with face $l$ \footnote{As we explained, it is straightforward to incorporate
fields in the adjoint representation to the proof.}. Let us consider two edges in $e_i$ and
$e_j$ in $E_x$ that are consecutive as we move around face $l$. Then,
$\Delta_{li}/\Delta_{lj}=1$ or $-1$ provided $e_i$ and $e_j$ are separated by and odd or even
number of edges, respectively. Conversely, $\textrm{sign}^x(e_i)/\textrm{sign}^x(e_j)=1$ or $-1$
if they are separated by and even or odd number of edges. Hence, we have that
$\textrm{sign}^x(e_i)\Delta_{li}/\textrm{sign}^x(e_j)\Delta_{lj}=-1$, and thus $\sum_{e_i \in
E_x} \textrm{sign}^{x}(e_i) \Delta_{li}=0$.

With identical reasoning, it follows that
\beq
\left( Q_D \ G_h^T \right)_{l y}= \sum_{e_i \in E_y} \textrm{sign}^{y}(e_i) \Delta_{li}=0
\label{QD_hy}
\eeq

From \eref{QD_hx} and \eref{QD_hy}, we conclude that
\beq
Q_D \ G_h^T=0
\eeq
Hence, we have $Q \ G_h^T=0$ and we can identify
\beq
G_h \equiv G
\eeq

We have shown that the slopes of the perfect matchings are
the positions of the corresponding GLSM fields in the toric diagram, completing our proof of the Mathematical Dimer Conjecture.

Before closing this section we notice an interesting result that
was possible due the use of dimers. Equations \eref{linear_h_T1} and \eref{linear_h_T2}
give the positions of GLSM fields in the toric diagram directly as linear combinations
of rows of $KT$. Nothing like these expressions was clear from the Forward
Algorithm and shows, once again, how dimers manage to pick the natural
variables for computing the moduli space.

\section{Conclusions}

In this paper we have proved the Mathematical Dimer Conjecture. That is, we have explicitly shown that there is a one--to--one mapping between the GLSM fields that realize the moduli space of a toric quiver and perfect matchings in the brane tiling dual to the periodic quiver. We have also demonstrated that the position of each GLSM field in the toric diagram is given by the slope of the corresponding perfect matching.

We have witnessed how dimers often provide an intuitive interpretation of otherwise obscure
steps in the computation of the moduli space. An example of this type is that F--term equations
can be easily solved using gauge transformations of weights as shown in Section
\ref{section_Fterm_gauge}. This leads to the magnetic flux variables and $\mathcal{V}$ as
natural intermediate variables of the Forward Algorithm.

There are several interesting directions that deserve further investigation. A partial list of them is:

\begin{itemize}

\item Our discussion has been limited to toric phases of the gauge theories (i.e. phases
in which all the gauge groups have the same rank). Non--toric phases are obtained by performing
a Seiberg duality transformation on a node for which the number of flavors is larger than twice
the number of colors. It would be interesting to investigate whether some generalization of the
brane tiling methods is applicable to these phases.

\item Conformal invariance can be broken by incorporating fractional branes (D5--branes wrapped over vanishing 2--cycles in the singular geometry). They modify the ranks of gauge groups in the quiver in a way that can be visualized in the brane tiling as a "chessboard" configuration \cite{Franco:2005zu}. The resulting RG flows take the form of duality cascades. It would be worth studying whether such RG flows are captured by some modification of the tiling.

\item Recently, there has been a renewed interest in marginal deformations of gauge
theories \cite{Leigh:1995ep} and the construction of their supergravity duals \cite{Lunin:2005jy}. Given the simplicity
with which superpotentials are encoded by brane tilings, it is natural to ask whether
and how it is possible to study this problem within this framework.

\item It is interesting to explore whether brane dimer methods can be extended to D($9-2p$)--branes probing
p--complex dimensional toric singularities. It is natural to conjecture that the corresponding
tilings will be ($p-1$)--dimensional and live on a ($p-1$)--dimensional torus.
The concepts of height function, slopes and characteristic polynomial should be appropriately
generalized to (p$-1$) dimensions. In analogy to what happens in four dimensions, if these constructions exist in other dimensions, they might be useful for finding possible field theory dualities.


\item Another direction is to investigate what is the geometric and
gauge theory meaning of brane tilings on the Klein Bottle, such as the one presented in Section \ref{section_conformal_invariance}.

\end{itemize}


\vskip 0.5cm

{\bf Acknowledgements:}
We gratefully acknowledge the invaluable discussions we have had with A.~Hanany, J.~Heckman, C.~Herzog, A.~King and C.~Vafa. The work of S. ~F. is supported by the National Science Foundation Grant No. PHY-0243680. D.~V. is supported in part by the CTP and the LNS of MIT and the U.S. Department of Energy under cooperative agreement $\#$DE-FC02-94ER40818.

\section{Appendix}
\section*{Perfect matchings for $dP_2$}

\begin{figure}[ht]
  \epsfxsize = 14cm
  \centerline{\epsfbox{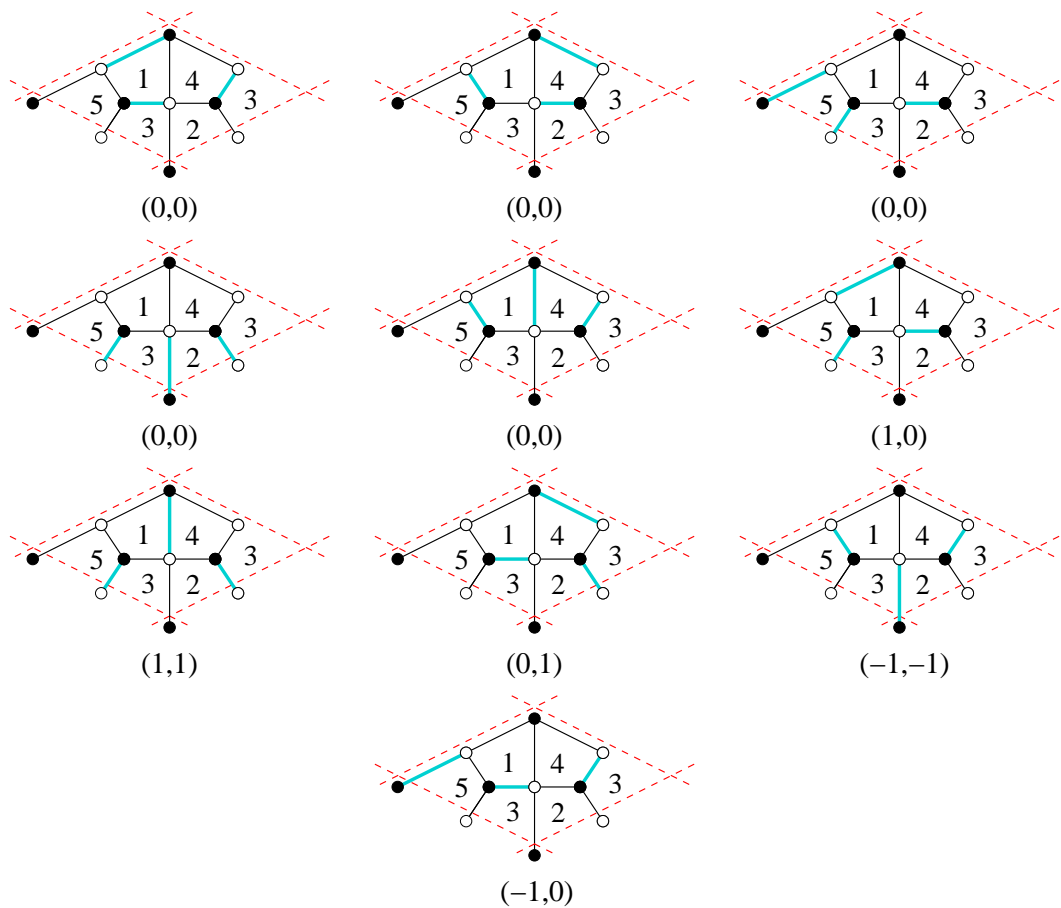}}
  \caption{Perfect matchings and their slopes for Model II of $dP_2$.}
  \label{perfect_matchings_dP2_II}
\end{figure}

\fref{perfect_matchings_dP2_II} presents the ten perfect matchings for Model II of $dP_2$ and their slopes.

\newpage

\bibliography{paper}

\providecommand{\href}[2]{#2}\begingroup\raggedright\begin{thebibliography}{10}

\bibitem{Maldacena:1997re}
J.~M. Maldacena, {\it The large n limit of superconformal field theories and
  supergravity},  {\em Adv. Theor. Math. Phys.} {\bf 2} (1998) 231--252,
  [\href{http://xxx.lanl.gov/abs/hep-th/9711200}{{\tt hep-th/9711200}}].

\bibitem{Gubser:1998bc}
S.~S. Gubser, I.~R. Klebanov, and A.~M. Polyakov, {\it Gauge theory correlators
  from non-critical string theory},  {\em Phys. Lett.} {\bf B428} (1998)
  105--114, [\href{http://xxx.lanl.gov/abs/hep-th/9802109}{{\tt
  hep-th/9802109}}].

\bibitem{Witten:1998qj}
E.~Witten, {\it Anti-de sitter space and holography},  {\em Adv. Theor. Math.
  Phys.} {\bf 2} (1998) 253--291,
  [\href{http://xxx.lanl.gov/abs/hep-th/9802150}{{\tt hep-th/9802150}}].

\bibitem{Aharony:1999ti}
O.~Aharony, S.~S. Gubser, J.~M. Maldacena, H.~Ooguri, and Y.~Oz, {\it Large n
  field theories, string theory and gravity},  {\em Phys. Rept.} {\bf 323}
  (2000) 183--386, [\href{http://xxx.lanl.gov/abs/hep-th/9905111}{{\tt
  hep-th/9905111}}].

\bibitem{Douglas:1996sw}
M.~R. Douglas and G.~W. Moore, {\it D-branes, quivers, and ale instantons},
  \href{http://xxx.lanl.gov/abs/hep-th/9603167}{{\tt hep-th/9603167}}.

\bibitem{Douglas:1997de}
M.~R. Douglas, B.~R. Greene, and D.~R. Morrison, {\it Orbifold resolution by
  d-branes},  {\em Nucl. Phys.} {\bf B506} (1997) 84--106,
  [\href{http://xxx.lanl.gov/abs/hep-th/9704151}{{\tt hep-th/9704151}}].

\bibitem{Intriligator:2003jj}
K.~Intriligator and B.~Wecht, {\it The exact superconformal r-symmetry
  maximizes a},  {\em Nucl. Phys.} {\bf B667} (2003) 183--200,
  [\href{http://xxx.lanl.gov/abs/hep-th/0304128}{{\tt hep-th/0304128}}].

\bibitem{Gauntlett:2004yd}
J.~P. Gauntlett, D.~Martelli, J.~Sparks, and D.~Waldram, {\it Sasaki-einstein
  metrics on s(2) x s(3)},  \href{http://xxx.lanl.gov/abs/hep-th/0403002}{{\tt
  hep-th/0403002}}.

\bibitem{Gauntlett:2004hh}
J.~P. Gauntlett, D.~Martelli, J.~F. Sparks, and D.~Waldram, {\it A new infinite
  class of sasaki-einstein manifolds},
  \href{http://xxx.lanl.gov/abs/hep-th/0403038}{{\tt hep-th/0403038}}.

\bibitem{Martelli:2004wu}
D.~Martelli and J.~Sparks, {\it Toric geometry, sasaki-einstein manifolds and a
  new infinite class of ads/cft duals},
  \href{http://xxx.lanl.gov/abs/hep-th/0411238}{{\tt hep-th/0411238}}.

\bibitem{Benvenuti:2004dy}
S.~Benvenuti, S.~Franco, A.~Hanany, D.~Martelli, and J.~Sparks, {\it An
  infinite family of superconformal quiver gauge theories with
  {S}asaki-{E}instein duals},
  \href{http://xxx.lanl.gov/abs/hep-th/0411264}{{\tt hep-th/0411264}}.

\bibitem{Cvetic:2005ft}
M.~Cvetic, H.~Lu, D.~N. Page, and C.~N. Pope, {\it New einstein-sasaki spaces
  in five and higher dimensions},
  \href{http://xxx.lanl.gov/abs/hep-th/0504225}{{\tt hep-th/0504225}}.

\bibitem{Martelli:2005wy}
D.~Martelli and J.~Sparks, {\it Toric sasaki-einstein metrics on s**2 x s**3},
  {\em Phys. Lett.} {\bf B621} (2005) 208--212,
  [\href{http://xxx.lanl.gov/abs/hep-th/0505027}{{\tt hep-th/0505027}}].

\bibitem{Cvetic:2005vk}
M.~Cvetic, H.~Lu, D.~N. Page, and C.~N. Pope, {\it New einstein-sasaki and
  einstein spaces from kerr-de sitter},
  \href{http://xxx.lanl.gov/abs/hep-th/0505223}{{\tt hep-th/0505223}}.

\bibitem{Franco:2005sm}
S.~Franco, A.~Hanany, D.~Martelli, J.~Sparks, D.~Vegh, and B.~Wecht, {\it Gauge
  theories from toric geometry and brane tilings},
  \href{http://xxx.lanl.gov/abs/hep-th/0505211}{{\tt hep-th/0505211}}.

\bibitem{Benvenuti:2005ja}
S.~Benvenuti and M.~Kruczenski, {\it From sasaki-einstein spaces to quivers via
  bps geodesics: Lpqr},  \href{http://xxx.lanl.gov/abs/hep-th/0505206}{{\tt
  hep-th/0505206}}.

\bibitem{Butti:2005sw}
A.~Butti, D.~Forcella, and A.~Zaffaroni, {\it The dual superconformal theory
  for lpqr manifolds},  \href{http://xxx.lanl.gov/abs/hep-th/0505220}{{\tt
  hep-th/0505220}}.

\bibitem{Martelli:2005tp}
D.~Martelli, J.~Sparks, and S.~T. Yau, {\it The geometric dual of
  $a$-maximisation for toric {S}asaki- {E}instein manifolds},
  \href{http://xxx.lanl.gov/abs/hep-th/0503183}{{\tt hep-th/0503183}}.

\bibitem{Butti:2005vn}
A.~Butti and A.~Zaffaroni, {\it R-charges from toric diagrams and the
  equivalence of a- maximization and z-minimization},  {\em JHEP} {\bf 11}
  (2005) 019, [\href{http://xxx.lanl.gov/abs/hep-th/0506232}{{\tt
  hep-th/0506232}}].

\bibitem{Tachikawa:2005tq}
Y.~Tachikawa, {\it Five-dimensional supergravity dual of a-maximization},
  \href{http://xxx.lanl.gov/abs/hep-th/0507057}{{\tt hep-th/0507057}}.

\bibitem{Barnes:2005bw}
E.~Barnes, E.~Gorbatov, K.~Intriligator, and J.~Wright, {\it Current
  correlators and ads/cft geometry},
  \href{http://xxx.lanl.gov/abs/hep-th/0507146}{{\tt hep-th/0507146}}.

\bibitem{Morrison:1998cs}
D.~R. Morrison and M.~R. Plesser, {\it Non-spherical horizons. i},  {\em Adv.
  Theor. Math. Phys.} {\bf 3} (1999) 1--81,
  [\href{http://xxx.lanl.gov/abs/hep-th/9810201}{{\tt hep-th/9810201}}].

\bibitem{Beasley:1999uz}
C.~Beasley, B.~R. Greene, C.~I. Lazaroiu, and M.~R. Plesser, {\it D3-branes on
  partial resolutions of abelian quotient singularities of calabi-yau
  threefolds},  {\em Nucl. Phys.} {\bf B566} (2000) 599--640,
  [\href{http://xxx.lanl.gov/abs/hep-th/9907186}{{\tt hep-th/9907186}}].

\bibitem{Feng:2000mi}
B.~Feng, A.~Hanany, and Y.-H. He, {\it D-brane gauge theories from toric
  singularities and toric duality},  {\em Nucl. Phys.} {\bf B595} (2001)
  165--200, [\href{http://xxx.lanl.gov/abs/hep-th/0003085}{{\tt
  hep-th/0003085}}].

\bibitem{Cachazo:2001sg}
F.~Cachazo, B.~Fiol, K.~A. Intriligator, S.~Katz, and C.~Vafa, {\it A geometric
  unification of dualities},  {\em Nucl. Phys.} {\bf B628} (2002) 3--78,
  [\href{http://xxx.lanl.gov/abs/hep-th/0110028}{{\tt hep-th/0110028}}].

\bibitem{Wijnholt:2002qz}
M.~Wijnholt, {\it Large volume perspective on branes at singularities},  {\em
  Adv. Theor. Math. Phys.} {\bf 7} (2004) 1117--1153,
  [\href{http://xxx.lanl.gov/abs/hep-th/0212021}{{\tt hep-th/0212021}}].

\bibitem{Herzog:2003dj}
C.~P. Herzog and J.~Walcher, {\it Dibaryons from exceptional collections},
  {\em JHEP} {\bf 09} (2003) 060,
  [\href{http://xxx.lanl.gov/abs/hep-th/0306298}{{\tt hep-th/0306298}}].

\bibitem{Herzog:2003zc}
C.~P. Herzog, {\it Exceptional collections and del pezzo gauge theories},  {\em
  JHEP} {\bf 04} (2004) 069,
  [\href{http://xxx.lanl.gov/abs/hep-th/0310262}{{\tt hep-th/0310262}}].

\bibitem{Aspinwall:2004bs}
P.~S. Aspinwall and S.~Katz, {\it Computation of superpotentials for d-branes},
   \href{http://xxx.lanl.gov/abs/hep-th/0412209}{{\tt hep-th/0412209}}.

\bibitem{Aspinwall:2005ur}
P.~S. Aspinwall and L.~M. Fidkowski, {\it Superpotentials for quiver gauge
  theories},  \href{http://xxx.lanl.gov/abs/hep-th/0506041}{{\tt
  hep-th/0506041}}.

\bibitem{Herzog:2004qw}
C.~P. Herzog, {\it Seiberg duality is an exceptional mutation},  {\em JHEP}
  {\bf 08} (2004) 064, [\href{http://xxx.lanl.gov/abs/hep-th/0405118}{{\tt
  hep-th/0405118}}].

\bibitem{Herzog:2005sy}
C.~P. Herzog and R.~L. Karp, {\it Exceptional collections and d-branes probing
  toric singularities},  \href{http://xxx.lanl.gov/abs/hep-th/0507175}{{\tt
  hep-th/0507175}}.

\bibitem{Hanany:2005ve}
A.~Hanany and K.~D. Kennaway, {\it Dimer models and toric diagrams},
  \href{http://xxx.lanl.gov/abs/hep-th/0503149}{{\tt hep-th/0503149}}.

\bibitem{Franco:2005rj}
S.~Franco, A.~Hanany, K.~D. Kennaway, D.~Vegh, and B.~Wecht, {\it Brane dimers
  and quiver gauge theories},
  \href{http://xxx.lanl.gov/abs/hep-th/0504110}{{\tt hep-th/0504110}}.

\bibitem{Hanany:2005ss}
A.~Hanany and D.~Vegh, {\it Quivers, tilings, branes and rhombi},
  \href{http://xxx.lanl.gov/abs/hep-th/0511063}{{\tt hep-th/0511063}}.

\bibitem{Feng:2005gw}
B.~Feng, Y.-H. He, K.~D. Kennaway, and C.~Vafa, {\it Dimer models from mirror
  symmetry and quivering amoebae},
  \href{http://xxx.lanl.gov/abs/hep-th/0511287}{{\tt hep-th/0511287}}.

\bibitem{Butti:2005ps}
A.~Butti and A.~Zaffaroni, {\it From toric geometry to quiver gauge theory: The
  equivalence of a-maximization and z-minimization},
  \href{http://xxx.lanl.gov/abs/hep-th/0512240}{{\tt hep-th/0512240}}.

\bibitem{Seiberg:1994pq}
N.~Seiberg, {\it Electric - magnetic duality in supersymmetric nonabelian gauge
  theories},  {\em Nucl. Phys.} {\bf B435} (1995) 129--146,
  [\href{http://xxx.lanl.gov/abs/hep-th/9411149}{{\tt hep-th/9411149}}].

\bibitem{Beasley:2001zp}
C.~E. Beasley and M.~R. Plesser, {\it Toric duality is {S}eiberg duality},
  {\em JHEP} {\bf 12} (2001) 001,
  [\href{http://xxx.lanl.gov/abs/hep-th/0109053}{{\tt hep-th/0109053}}].

\bibitem{Feng:2001bn}
B.~Feng, A.~Hanany, Y.-H. He, and A.~M. Uranga, {\it Toric duality as {S}eiberg
  duality and brane diamonds},  {\em JHEP} {\bf 12} (2001) 035,
  [\href{http://xxx.lanl.gov/abs/hep-th/0109063}{{\tt hep-th/0109063}}].

\bibitem{Franco:2003ja}
S.~Franco, A.~Hanany, Y.-H. He, and P.~Kazakopoulos, {\it Duality walls,
  duality trees and fractional branes},
  \href{http://xxx.lanl.gov/abs/hep-th/0306092}{{\tt hep-th/0306092}}.

\bibitem{Feng:2002zw}
B.~Feng, S.~Franco, A.~Hanany, and Y.-H. He, {\it Symmetries of toric duality},
   {\em JHEP} {\bf 12} (2002) 076,
  [\href{http://xxx.lanl.gov/abs/hep-th/0205144}{{\tt hep-th/0205144}}].

\bibitem{Benvenuti:2005cz}
S.~Benvenuti and M.~Kruczenski, {\it Semiclassical strings in sasaki-einstein
  manifolds and long operators in n = 1 gauge theories},
  \href{http://xxx.lanl.gov/abs/hep-th/0505046}{{\tt hep-th/0505046}}.

\bibitem{Hanany:1997tb}
A.~Hanany and A.~Zaffaroni, {\it On the realization of chiral four-dimensional
  gauge theories using branes},  {\em JHEP} {\bf 05} (1998) 001,
  [\href{http://xxx.lanl.gov/abs/hep-th/9801134}{{\tt hep-th/9801134}}].

\bibitem{Hanany:1998it}
A.~Hanany and A.~M. Uranga, {\it Brane boxes and branes on singularities},
  {\em JHEP} {\bf 05} (1998) 013,
  [\href{http://xxx.lanl.gov/abs/hep-th/9805139}{{\tt hep-th/9805139}}].

\bibitem{Aganagic:1999fe}
M.~Aganagic, A.~Karch, D.~Lust, and A.~Miemiec, {\it Mirror symmetries for
  brane configurations and branes at singularities},  {\em Nucl. Phys.} {\bf
  B569} (2000) 277--302, [\href{http://xxx.lanl.gov/abs/hep-th/9903093}{{\tt
  hep-th/9903093}}].

\bibitem{bondy}
J.~A. Bondy and U.~S.~R. Murty, {\em Graph Theory with Applications}.
\newblock North-Holland, 1976.

\bibitem{Kenyon:2002a}
R.~Kenyon, {\it An introduction to the dimer model},
  \href{http://xxx.lanl.gov/abs/math.CO/0310326}{{\tt math.CO/0310326}}.

\bibitem{Kenyon:2003uj}
R.~Kenyon, A.~Okounkov, and S.~Sheffield, {\it Dimers and amoebae},
  \href{http://xxx.lanl.gov/abs/math-ph/0311005}{{\tt math-ph/0311005}}.

\bibitem{Edmonds:1965}
J.~Edmonds, {\it Maximum matching and a polyhedron with (0,1)-vertices},  {\em
  J. Res. Nat. Bur. Standards} {\bf B69} (1965) 125--130.

\bibitem{Franco:2005zu}
S.~Franco, A.~Hanany, F.~Saad, and A.~M. Uranga, {\it Fractional branes and
  dynamical supersymmetry breaking},
  \href{http://xxx.lanl.gov/abs/hep-th/0505040}{{\tt hep-th/0505040}}.

\bibitem{Leigh:1995ep}
R.~G. Leigh and M.~J. Strassler, {\it Exactly marginal operators and duality in
  four-dimensional n=1 supersymmetric gauge theory},  {\em Nucl. Phys.} {\bf
  B447} (1995) 95--136, [\href{http://xxx.lanl.gov/abs/hep-th/9503121}{{\tt
  hep-th/9503121}}].

\bibitem{Lunin:2005jy}
O.~Lunin and J.~Maldacena, {\it Deforming field theories with u(1) x u(1)
  global symmetry and their gravity duals},  {\em JHEP} {\bf 05} (2005) 033,
  [\href{http://xxx.lanl.gov/abs/hep-th/0502086}{{\tt hep-th/0502086}}].

\end{thebibliography}\endgroup
\bibliographystyle{JHEP}

\end{document}